\documentclass[preprint,12pt]{elsarticle}
\usepackage{amssymb}
\usepackage{amsthm}
\usepackage{amsmath}
\usepackage{pbox}
\usepackage{booktabs}
\usepackage[usenames, dvipsnames]{color}
\usepackage{nicefrac}
\usepackage{lineno}
\usepackage{float}
\usepackage{hyperref}
\usepackage{pdflscape}
\usepackage{rotating}
\journal{Journal of Biological Physics}
\begin{document}
% \linenumbers
\begin{frontmatter}
	\title{Fitting of dynamic recurrent neural network models to sensory stimulus-response data}
	
	
	\author[jhmi1,atlm]{R.Ozgur DORUK\corref{cor1}\fnref{grant1}}
	\cortext[cor1]{Corresponding Author}
	\fntext[label2]{This work is partially supported by Turkish Scientific and Technological Research Council (T\"{U}B\.{I}TAK) 2219 Research Program.}
	\fntext[label3]{The computational facilities needed in this research are provided by High Performance Computing (HPC) center TRUBA/TR-GRID owned by National Academic Information Center (ULAKBIM) of TURKEY}
	\ead{resat.doruk@atilim.edu.tr}
	\author[jhmi2]{Kechen Zhang}
	% \ead{kzhang4@jhmi.edu}
	\address[atlm]{Atilim University, Department of Electrical and Electronics Engineering, Kizilcasar Mahallesi, Incek, Golbasi, Ankara, 06836,TURKEY}
	\address[jhmi1]{Department of Biomedical Engineering, Johns Hopkins School of Medicine, 720 Rutland Avenue, Ross 528, Baltimore , MD, 21205, USA}
	\address[jhmi2]{Department of Biomedical Engineering, Johns Hopkins School of Medicine, 720 Rutland Avenue, Traylor 407, Baltimore , MD, 21205, USA}
	\begin{abstract}
		We present a theoretical study aiming at model fitting for sensory neurons. Conventional neural network training approaches are not applicable to this problem due to lack of continuous data. Although the stimulus can be considered as a smooth time dependent variable, the associated response will be a set of neural spike timings (roughly the instants of successive action potential peaks) which have no amplitude information. A recurrent neural network model can be fitted to such a stimulus-response data pair by using maximum likelihood estimation method where the likelihood function is derived from Poisson statistics of neural spiking. The universal approximation feature of the recurrent dynamical neuron network models allow us to describe excitatory-inhibitory characteristics of an actual sensory neural network with any desired number of neurons. The stimulus data is generated by a Phased Cosine Fourier series having fixed amplitude and frequency but a randomly shot phase. Various values of amplitude, stimulus component size and sample size are applied in order to examine the effect of stimulus to the identification process. Results are presented in tabular form at the end of this text.  
		
	\end{abstract}
\end{frontmatter}

\section{Introduction}
\label{sec:intro}
\subsection{General Discussion on Neurons and Information Flow}\label{sub:intro-gen-discuss}
Theoretical or computational neuroscience is a recent field of research emerged after the development of mathematical models of real biological neurons. The Hodgkin-Huxley model \cite{hodgkin1952quantitative} which can be considered as a biological oscillator is the first sounding attempt in this field. After that, a lot of similar research is conducted and simpler or more complicated models are derived. Most of these involve the membrane potential as the main dynamical variable (Fitzhugh-Nagumo \cite{fitzhugh1961impulses}, Morris-Lecar \cite{morris1981voltage} models). On the other hand some others involve different variables. One example that seemed to have a crucial position in computational neuro-science is the neural firing rate based model \cite{ledoux2011dynamics} which is actually an extension to the continuous time dynamical recurrent neural network \cite{beer1995dynamics,miller2012mathematical}. The efficiency and usability of these models depend on the aim of the research and the limitations set by the simulation/experiment environment. In experiments related to computational neuroscience field, one such limitation may arise from the measurement capability. In vivo experiments, does not allow the real time measurement of the membrane potential. An attempt to achieve this will likely to interrupt the propagation of action potentials due to a change in the axial membrane physical properties at the instant of electrode placement. In some cases the neuron may be damaged. Thus, the most practical way to gather data in vivo from a live neuron is to record the instants of successive action potentials (in other words the spiking instants) using an electrode attached at a site in the surrounding medium. By that, the current flow through the surrounding conductance helps us to record the spiking times. Concerning theoretical or computational neuro-science studies that will be quite interesting. Recent studies such as \cite{gerstner1997neural} suggests that the information transmitted along the sensory and motor neurons are coded somehow by the temporal locations of the spikes and/or the associated firing rates. So the timings of the spikes can be collected by placing an electrode in the surroundings of the studied neuron. Another tackling feature of the neural spiking phenomena is that, it is not a deterministic event. The stochasticity of the ion channels \cite{herz2006modeling} and synaptic noise led to the fact that the data transmitted along the neurons is corrupted by noise. Again from related research \cite{shadlen1994noise} it can be noted that, this stochasticity of neural spiking obeys the famous Inhomogeneous Poisson Process at least for the sensory neurons. So a proper likelihood methodology may aid the parameter identification procedures. 
\subsection{Modeling}\label{sub:intro-modeling}
Knowing the fact that there are a dozen of neuron models in the literature, a question will arise: How type of a model should we use?. In this research our aim is to identify the parameters of a neuron model based on the recorded stimulus-response data. As the response data does not reflect any membrane potential information but the distribution of the neural spikes instead, a model reflecting the firing rate will be much meaningful for this research. So, one may eliminate the complicated models like Hodgkin-Huxley or Morris Lecar. Instead, we may use a more generic model where the number of neurons can be set to any desired value. Based on these facts, a continuous time generic dynamical recurrent neural network (CTRNN) model can fit this purpose. CTRNNs can be modeled in two forms. One employs the membrane potential variable as its states (but no channel related dynamics explicitly modeled, they are embedded into model) and the other presents the dynamics of the neural firing rates directly as states. The former can provide the firing rate as an output variable. The two types are proven to be equivalent \cite{miller2012mathematical}. In this research we prefer the first one, namely the membrane potential based one and the firing rate will be mapped through a sigmoidal function. See \textbf{Section \ref{sub:ctrnn-theory}} for details. In addition, some of the neurons in the selected CTRNN can be made excitatory and others be inhibitory. Doing this will allow one to model the firing and refractory response of the neuron more truly.  The dynamic properties of the neuron membrane is represented by time constants and the synaptic excitation and inhibition are represented as network weights (scalar gains). Though not the same, a similar excitatory-inhibitory structure is utilized in numerous studies such as \cite{hancock1997modeling,hancock1999wideband,de2008linking}.\subsection{Parameter Identification}
\label{sub:intro-par-id}
Having chosen the model structure, one has to decide how the parameters will be estimated. The first discussion is centered around the structure of the stimulus driving the neural network. There can be various forms for stimulus. As the study targets auditory cortex, a sine related stimulus can be chosen where a stimulus modeled by a Fourier Series seemed to be a good choice. 

Concerning parameter estimation, the best choice is to develop a likelihood based approach as one can only talk about the statistics of the collected spike times. As the timing is stochastic and supposed to obey the Inhomogeneous Poisson statistics we can employ a maximum likelihood estimation procedure. The likelihood function will be derived from the Inhomogeneous Poisson probability mass function or a more specific one developed by \cite{uteden2008pointprocesses,brown2002time}. The latter one is expected to provide a better identification result. The reason for this is that, the second likelihood is a function of firing rate and individual spike times where as the former one only requires the number of spikes other than the firing rate. So the firing rate output of identified CTRNN is expected to approach to the true firing rate as the identification algorithm knows the location of the spikes. 
\subsection{Challenges}
\label{sub:intro-challng}
There are certain challenges in this research. First of all, we will most probably not be able to have a reasonable estimate just from a single spiking response data set as we do not have a continuous response data. This is also demonstrated in the related kernel density estimation research such as \cite{nawrot1999single,shimazaki2010kernel,shimazaki2007method,koyama2004histogram}. From these sources, one will easily note that repeated trials and superimposed spike sequences are required to obtain a meaningfully accurate firing rate information from the neural response data. In a real experiment environment, repeating the trials with the same stimulus profile will not be appropriate as the repeated responses of the same stimulus are found to be attenuated.
Because of this issue, a new stimulus should be provided at each excitation. This can be provided by choosing a fixed amplitude and frequency but randomly shot phase angle for our Fourier Series stimulus. 
Secondly, in the likelihood estimation, the complete data from the beginning will be used in the likelihood optimization. This will be a computational challenge as a very large data will be accumulated in each computation step. When considering an experiment we collect the data only by providing a random stimulus entry to the animal (experiment subject) and record the spike counts and locations. As animal is not involved in the computational part of the random stimuli based experiments an high performance computing (HPC) facility can be involved without a need of any wet experimental element. In this research, we are employing the high performance computing facilities (TRUBA/TR-GRID) of the National Academic Information Center (ULAKBIM) of Turkish Scientific and Technological Research Institution (TUBITAK). 
\subsection{Previous Studies}
\label{sub:intro-prev-stud} 
This work is a fairly novel attempt. There are very few studies in the literature that have a similar goal. Some examples can be given as \cite{brillinger1988maximum,brown2002time,chornoboy1988maximum,paninski2004maximum,smith2003estimating}. \textcolor{BrickRed}{The work in \cite{brillinger1988maximum} presents a system identification study based on maximum likelihood estimation of the internal parameters of an integrate and fire neuron model. Likelihood function is derived from firing probabilities through local Bernoulli approximation. \cite{chornoboy1988maximum} aims at the detection of the functional relationships between neurons. Rather than modeling an individual neuron, it involves a characterization of the neural interactions through maximum likelihood estimation. \cite{paninski2004maximum} is somehow similar to \cite{brillinger1988maximum}. A thorough explanation of maximum likelihood explanation is presented with an application to a linear-nonlinear Poisson cascade and an integrate and fire model generalized linear model. It also presents a comparison with traditional spike triggered average estimator. \cite{smith2003estimating} presents a similar work to that of \cite{brillinger1988maximum} and \cite{paninski2004maximum} with a different model. The model involve an estimation of a conditional intensity function modulated by an unobservable latent state-space process. Study also involves the identification of the latent process. Both estimation approaches are based on maximum likelihood method. \cite{paninski2004maximum} and \cite{smith2003estimating} applies expectation maximization method in the solution of the maximum likelihood problems. For a more general discussion on the application of statistical techniques and their challenges in theoretical and computational neuroscience interested readers can apply to the reference \cite{wu2006complete}.}   

\textcolor{Brown}{This research has some common grounds with \cite{brillinger1988maximum} and \cite{paninski2004maximum} due to the application of maximum likelihood method to a neural network identification problem. However the model used in this research is quite different from the ones in those sources. Instead of an integrate and fire model we prefer a more general continuous time recurrent neural network due to their universal approximation capability which is expected to be an advantage to model a multi-cellular region of the nervous system. In addition their dynamical properties are closer to network models such as Hodgkin-Huxley or Moris-Lecar equations. Research such as \cite{dimattina2011active,dimattina2013adaptive} implements a generic neural network model which is of the a static feed-forward type. Based on all these, one can say that this study can be considered as a novel contribution to the neuroscience literature. In addition the work done in \cite{brillinger1988maximum,brown2002time,chornoboy1988maximum,paninski2004maximum,smith2003estimating} are too elaborate in statistical theory with a very limited discussion on how to apply the theory to neuron modeling. This restricts the reproducibility of those research. This text concentrates also on how to apply the theory on the identification problem using computational tools such as MATLAB to increase its reproducibility.}

\section{Models \& Methods}
\label{sec:models-methods}

\subsection{Continuous Time Recurrent Neural Networks (CTRNN)} \label{sub:ctrnn-theory}
\begin{figure}[H]
	\centering
	\includegraphics[scale=0.5]{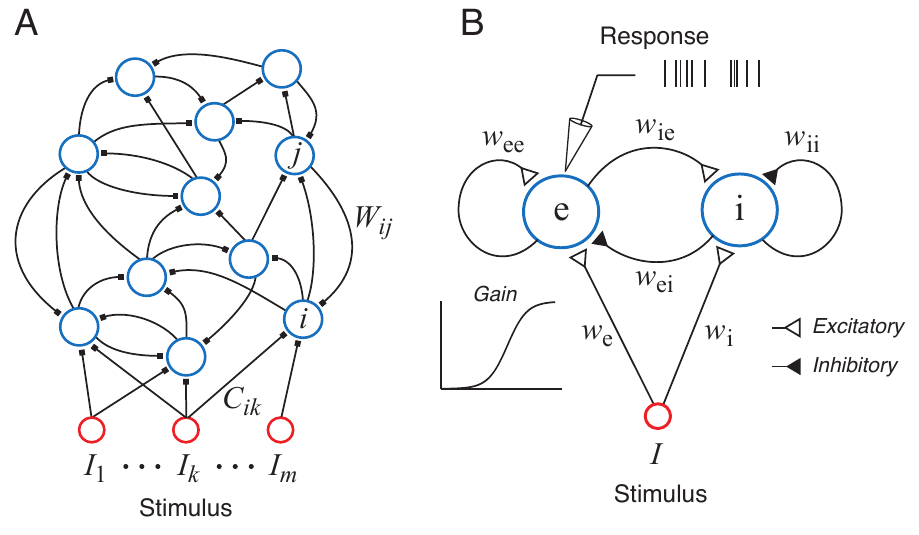}
	\caption{(\textbf{A}) A generic recurrent neural network structure. The stimulus means external inputs to the network.
		(\textbf{B}) A simple recurrent network with one excitatory unit and one inhibitory unit, with both units having nonlinear sigmoidal gain functions. Here each unit may represent a population of neurons.We assume that the recorded responses are inhomogeneous Poisson spike trains based on the continuous rate generated by the state of the excitatory unit.  }
	\label{fig:generic-ctrnn}
\end{figure}

The continuous time recurrent neural networks have a similar structure to that of the discrete time counterparts that are often met in artificial intelligence studies. In \textbf{Figure \ref{fig:generic-ctrnn}}, one can see a general continuous time network that may have any number of neurons.
The mathematical representation of this generic model can be written as shown below \cite{beer1995dynamics}:      
\begin{equation}
\tau_{i}\frac{dV_{i}}{dt}=-V_{i}+\sum_{j=1}^{n} W_{ij}g_{j}\left(V_{j}\right) +\sum_{k=1}^{m} C_{ik}I_{k}
\label{eq:general-network-model}
\end{equation}
\noindent where ${\tau_i}$ is the time constant, ${V_i}$ is the membrane potential of the ${i^{th}}$ neuron, ${W_{ij}}$ is the synaptic connection weight between the ${i^{th}}$ and ${j^{th}}$ neurons ${C_{ik}}$ is the connection weight from ${i^{th}}$ input to the ${k^{th}}$ neuron and ${I_k}$ is the ${k^{th}}$ input. The term ${g_j\left(V_j\right)}$ is a membrane potential dependent function which acts as a variable gain on the synaptic inputs to from the ${j^{th}}$ neuron to the ${k^{th}}$ one. It can be shown by a logistic sigmoid function which can be shown as:
\begin{equation}
g_{j}\left(V_{j}\right)=\frac{\Gamma_j}{1+\exp\left( -a_{j}\left(V_{j}-h_j\right)\right) }\label{eq:sigmoid-general}
\end{equation} 
where $\Gamma_j$ is the maximum rate at which the ${j^{th}}$ neuron can fire, $h_j$ is a soft threshold parameter of the ${j^{th}}$ neuron and ${a_j}$ is a slope constant. This is the only source of non-linearity in \eqref{eq:general-network-model}. \textcolor{OrangeRed}{Similar functional forms are also met in more complicated neuron models such as Hodgkin-Huxley equation \cite{hodgkin1952quantitative}. The equations describing the dynamics of the channel activations and inactivations involve sigmoid functions like \eqref{eq:sigmoid-general}}. The work by \citep{miller2012mathematical} shows that \eqref{eq:sigmoid-general} gives a relationship between the firing rate ${r_j}$ and membrane potential ${V_j}$ of the ${j^{th}}$ neuron. In sensory nervous system, some of neurons have excitatory synaptic connections while some have inhibitory ones. This fact is reflected to the model in \eqref{eq:general-network-model} by assigning negative values to the weight parameters which are originating from neurons with inhibitory synaptic connections. \textcolor{RubineRed}{At this point, one has to note that As shown in  a CTRNNs may have any number of neurons with multiple inputs, outputs and layers (see \textbf{Figure \ref{fig:generic-ctrnn}a}). Depending on the applications a complicated neural network may or may not be necessary. Regardless of that, having large numbers of neurons will increase constitute a computational burden. As we are desiring to prove the methodology presented in this text, it will be beneficial to start with a basic model. This should also be a right choice as there are a very few number of similar studies which will guide the researchers. So we choose a CTRNN with only two neurons. In this contect, we will assume that the dynamics od the excitatory and inhibitory members of a part of the auditory cortex are lumped into two neurons. Here one neuron will represent the average response of the excitatory neurons and be denoted by subscript $ ._e $ and the other will be denoted by subscript $ ._i $ and represent the average response of the inhibitory neurons in the network. The stimulus will also be represented by a single input signal and distributed to the neurons by weights. As stated this approach is preferred to validate the development in this research. Of course, it can be extended to a network with any number of neurons and layers (like \eqref{eq:general-network-model}). }So a basic excitatory and inhibitory continuous time recurrent dynamical network can be written as shown in the following:
\begin{eqnarray}
\tau_{e}\dot{V}_{e}&=&-V_{e}+w_{ee}g_{e}\left(V_{e}\right)-w_{ei}g_{i}\left(V_{i}\right)+w_{e}I
\label{eq:2-exc-inh-network-a}
%\nonumber 
\\
\tau_{i}\dot{V_{i}}&=&-V_{i}++w_{ie}g_{e}\left(V_{e}\right)-w_{ii}g_{i}\left(V_{i}\right)+w_{i}I
\label{eq:2-exc-inh-network-b}
\end{eqnarray}
where $ V_e $ and $ V_i $ are the membrane potentials of the individual excitatory and inhibitory neurons respectively. \textcolor{ForestGreen}{As we have just mentioned, the response of all excitatory and inhibitory units are lumped into two single neurons connecting to excitatory and inhibitory synapses respectively. Stimulus is represented by a single input that is ${I}$}. In addition in order to suit the model equations to the estimation theory formalism the time constant may be moved to the right hand side as shown below:
\begin{equation} 
\frac{d}{dt}\left[\begin{array}{c}
V_{e}\\
V_{i}
\end{array}\right]=\left[\begin{array}{cc}
\beta_{e} & 0\\
0 & \beta_{i}
\end{array}\right]\left\{ -\left[\begin{array}{c}
V_{e}\\
V_{i}
\end{array}\right]+\left[\begin{array}{cc}
w_{ee} & -w_{ei}\\
w_{ie} & -w_{ii}
\end{array}\right]\left[\begin{array}{c}
g_{e}\left(V_{e}\right)\\
g_{i}\left(V_{i}\right)
\end{array}\right]+\left[\begin{array}{c}
w_{e}\\
w_{i}
\end{array}\right]I\right\} \label{eq:our-model-matrix-form}
\end{equation}
where $ \beta_e $ and $ \beta_i $ are the reciprocals of the time constants $ \tau_e $ and $ \tau_i $. They are taken to the right for easier manipulations of the equations.	
\noindent Note that this equation is written in matrix form to be suit the formal non-linear system forms. A descriptive illustration related to \eqref{eq:our-model-matrix-form} is presented in \textbf{Figure \ref{fig:generic-ctrnn}b}. It should also be noted that, in \eqref{eq:2-exc-inh-network-b} and \eqref{eq:our-model-matrix-form} the weights are all assumed as positive coefficients and they have signs in the equation. So negative signs indicate that originating neuron is inhibitory (tend to hyper-polarize the other neurons in the network).  

\subsection{Inhomogeneous Poisson spike model}\label{sub:poisson-theory} 

The theoretical response of the network in \eqref{eq:2-exc-inh-network-b} will be the firing rate of the excitatory neuron as ${r_e=g_e\left(V_e\right)}$. In the actual environment, the neural spiking due to the firing rate ${r_e\left(t\right)}$ is available instead. While introducing this research, it is stated that this spiking events conform to an inhomogeneous Poisson process which is defined below:
\begin{equation}
\mbox{Prob}\left[N\left(t+\Delta t\right)-N\left(t\right)=k\right]=\frac{e^{-\lambda} \lambda^{k}}{k!}
\label{eq:inhomogeneous-poisson}
\end{equation} 
where
\begin{equation}
\lambda=\int_{t}^{t+\Delta t}r_e\left(\tau\right)d\tau
\end{equation}
is the mean number of spikes based on the firing rate $r_e(t)$ which varies with time,  and $N(\tau)$ indicates the cumulative total number of spikes up to time $\tau$, so that  $N\left(t+\Delta t\right)-N\left(t\right)$ is the number of spikes within the time interval ${\left[t,t+\Delta t\right)}$.

In other words, the probability of having ${k}$ number of spikes in the interval ${\left(t,t+\Delta t\right)}$ is given by the Poisson distribution above. 

Consider a spike train  $(t_1,t_2,\ldots,t_K)$ in the time interval $(0, T)$ (here $ 0\leq t_1 \leq t_2 \leq \ldots \leq t_K \leq T $ so $ t $ and $ \Delta t $ become $ t=0 $ and $ \Delta t=T $).  Here the spike train is described by a list of the time stamps for the $K$ spikes.  
The probability density function for a given spiking train $(t_1,t_2,\ldots,t_K)$ can be derived from the inhomogeneous Poisson process  \cite{uteden2008pointprocesses,brown2002time}. The result reads:
	\begin{equation}
	% p({\bf r} \mid {\bf x},\theta)
	p\left(t_{1},t_{2},\ldots,t_{K}\right)
	=\exp\left(-\int_{0}^{T}r_e\left(t,{\bf x},\theta\right)dt\right)\prod_{k=1}^K r_e\left(t_k,{\bf x},\theta\right)
	\label{eq:lkl-enbrown}
	\end{equation}  
This probability density describes how likely a particular spike train $(t_1,t_2,\ldots,t_K)$ is generated by the inhomogeneous Poisson process with the rate function $r_e\left(t,{\bf x},\theta\right)$., Of course, this rate function depends implicitly on the network parameters and the stimulus used.
\subsection{Maximum Likelihood Methods and Parameter Estimation} \label{sub:jmle-theory}

The network parameters to be estimated are listed below as a vector:
\begin{equation}
\theta=\left[\theta_1, \ldots,\theta_8\right]
=\left[\beta_e,\beta_i,w_e,w_i,w_{ee},w_{ei},w_{ie},w_{ii}\right]\label{eq:theta-ctrnn-param}
\end{equation}
which includes the time constants and all the connection weights in the E-I network.
Our maximum-likelihood estimation of the network parameters is based on the likelihood function given by \eqref{eq:lkl-enbrown}, 
which takes the individual spike timings into account. 
It is well known from estimation theory is that maximum likelihood estimation is asymptotically efficient, i.e., reaching the Cram\'er-Rao bound in the limit of large data size.
To extend the likelihood function in \eqref{eq:lkl-enbrown} to the situation where there are multiple spike trains elicited by multiple stimuli,
consider a sequence of $M$ stimuli. \textcolor{Violet}{This means that we drive the network in \eqref{eq:our-model-matrix-form} $ M $ times by generating $ M $ different stimuli at each trial. If $ I_j $ and $ I_k $ are the stimuli for the $ j^{th} $ and $ k^{th} $ trials respectively for $ j,k=1,\ldots,M $, $ I_j \neq I_k $ for all cases where $ j \neq k $.} 
Suppose the $m$-th stimulus ($m=1,\ldots, M$) elicits a spike train with a total of $K_m$ spikes in the time window $[0,T]$, and the spike timings are given by
${S}_m=\left(t_1^{(m)},t_2^{(m)},\ldots,t_{K_m}^{(m)}\right)$. 
By \eqref{eq:lkl-enbrown}, the likelihood function for the spike train $S_m$ is
\begin{equation}
p\left(S_m\mid\theta\right)
=\exp\left(-\int_{0}^{T}r_e^{(m)}\!\left(t\right)dt\right)\prod_{k=1}^{K_m} r_e^{(m)}\!\left(t_{k}^{(m)}\right)
\label{eq:pSm}
\end{equation}     
where $r_e^{(m)}$ is the firing rate in response to the $m$-th stimulus. Note that the rate function $r_e^{(m)}$  depends implicitly on the network parameters $\theta$ and on the stimulus parameters. The left-hand side of \eqref{eq:pSm} emphasizes the dependence on network parameters $\theta$, which is convenient for parameter estimation. 
The dependence on the stimulus parameters will be discussed in the next section.

We assume that the responses to different stimuli are independent, which is a reasonable assumption when the inter-stimulus intervals are sufficiently large. Under this assumption, the overall likelihood function for the collection of all $M$ spike trains can be written as
\begin{equation}
L\left({S}_{1},S_{2},\ldots,S_{M} \mid \theta \right)
=\prod_{m=1}^{M}p\left(S_m \mid\theta\right)
\label{eq:joint-likelihood-product}
\end{equation}
By taking natural logarithm, we obtain the log likelihood function:
\begin{equation}
l \left({S}_{1},S_{2},\ldots,S_{M} \mid \theta\right)
=-\sum_{m=1}^{M} \int_{0}^{T}r_e^{(m)}\!\left(t\right)dt 
+\sum_{m=1}^{M}\sum_{k=1}^{K_m}\ln{r_e^{(m)}}\!\left(t_{k}^{(m)}\right)
\label{eq:complete-likelihood-compact}
\end{equation}
Maximum-likelihood estimation of the parameter set is given formally by
\begin{equation}
\hat\theta_{ML}=\arg\max_{\theta} \left[ l  \left({S}_{1},S_{2},\ldots,S_{M} \mid \theta\right)\right] 
\label{eq:mle-arg-max-log}
\end{equation}
\subsection{Stimulus}
\label{sub:stimulus}
As discussed in \textbf{Section \ref{sub:intro-par-id}}, we will model the stimulus signal by a phased cosine Fourier series as shown below:
\begin{equation}
I=\sum_{n=1}^{N}A_n \cos\left(\omega_{n}t+\phi_{n}\right)\label{eq:cosine-stimulus}
\end{equation}
where $A_n$ is the amplitude, $\omega_n$ is the frequency of the $n$-th Fourier component, and $\phi_n$ is the phase of the component. Here the amplitude $A_n$ and frequency $\omega_n$ are fixed but the phase $\phi_n$ will be a randomly chosen from a uniform distribution between $ [-\pi,\ \pi] $ radians.        
\section{Results} \label{sub:results}
In this section, we will summarize the functional and numerical details of the neural network parameter estimation algorithm. 
\subsection{Details of the example model} \label{sub:example-details}
\begin{figure}[H]
	\centering
	\includegraphics[scale=0.5]{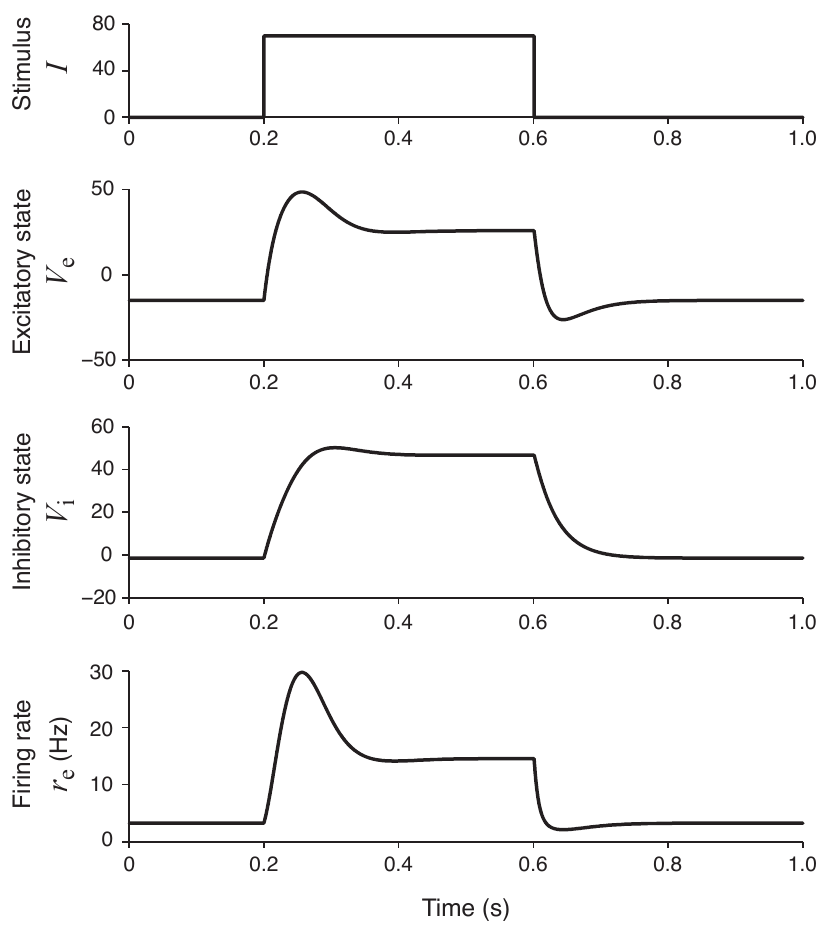}
	\caption{ The network model in \textbf{Figure \ref{fig:generic-ctrnn}B} in response to a square-wave stimulus. The states of the excitatory and inhibitory units, $V_e$ and $V_i$, are shown, together with the continuous firing rate of the excitatory unit, $r_e=g_e\left(V_e\right)$. The firing rate of the excitatory unit (bottom panel) has a transient component with higher firing rates, followed by a plateau or sustained component with lower firing rates.}
	\label{fig:pulse-response-ctrnn}
\end{figure}

This section is devoted to the detailed presentation of the simulation set-up. An numerical example will be presented which will demonstrate our approach. 
In the example application, the algorithms  presented in \textbf{Section \ref{sub:jmle-theory}} are applied to probe an EI network. 
\noindent In order to verify the performance of the parameter estimation we have to compare the estimates with their true values. So we will need a set of reference values of the model parameters in \eqref{eq:our-model-matrix-form}. These are shown in \textbf{Table \ref{tab:true-values-parameters}}. The example model can also be seen in \textbf{Figure \ref{fig:generic-ctrnn}b}. 
\begin{table}[H]
	\centering
	\caption{ 
		The true values of the parameters of the network model in \eqref{eq:our-model-matrix-form}. These are the parameters to be estimated. 
	} \label{tab:true-values-parameters}
	\begin{tabular}{ccc}
		\toprule 
		Parameter & Unit & True value $\left(\theta\right)$ \\
		\midrule
		${\beta_e}$ & $ \nicefrac{1}{s} $ & ${50}$ \\
		\midrule
		${\beta_i}$ & $ \nicefrac{1}{s} $ & ${25}$ \\
		\midrule
		${w_e}$ & k$ \Omega $ & ${1.0}$ \\
		\midrule
		${w_i}$ & k$ \Omega $ & ${0.7}$ \\
		\midrule
		${w_{ee}}$ & mV$ \cdot $s & ${1.2}$  \\
		\midrule
		${w_{ei}}$ & mV$ \cdot $s & ${2.0}$ \\
		\midrule
		${w_{ie}}$ & mV$ \cdot $s & ${0.7}$ \\
		\midrule 
		${w_{ii}}$ & mV$ \cdot $s & ${0.4}$ \\
		\bottomrule 
	\end{tabular}    
\end{table}
Our model in \eqref{eq:our-model-matrix-form} has two more important components which are the gain functions $g_e\left(V_e\right)$ and $g_i\left(V_i\right)$. These are obtained by setting $j$ in \eqref{eq:sigmoid-general} by either '$e$' or '$i$'. So one has $6$ additional parameters $\left[\Gamma_e,a_e,h_e,\Gamma_i,a_i,h_i\right]$ which have direct effect on the neural model behaviour. \textcolor{Blue}{This research targets the estimation of the network weights $ (w_e,w_i,w_{ee},w_{ei},w_{ie},w_{ii}) $ and reciprocal time constants $ (\beta_e,\beta_i) $ only}. Because of that, the parameters of the gain functions are assumed to be known and they have the values as shown in \textbf{Table \ref{tab:sigmoid-values-ge-gi}}. 
\begin{table}[H]
	\centering
	\caption{The parameters of the sigmoidal gain functions $ g_j(V) $ in \eqref{eq:sigmoid-general} for the excitatory (e) and inhibitory (i) neurons of the example model.}\label{tab:sigmoid-values-ge-gi}
	\begin{tabular}{cc}
		\toprule 
		Parameter & Value \\ 
		\midrule 
		$ \Gamma_e $ & 100 \\ 
		\midrule
		$ a_e $ & 0.04 \\
		\midrule
		$ h_e $ & 70 \\
		\midrule 
		$ \Gamma_i $ & 50 \\ 
		\midrule
		$ a_i $ & 0.04 \\
		\midrule
		$ h_i $ & 35 \\ 
		\bottomrule
	\end{tabular} 
\end{table} 
\noindent This set of parameters (gain functions and \textbf{Table \ref{tab:true-values-parameters}}) allows the network to have a unique equilibrium state for each stationary input. To demonstrate the excitatory and inhibitory characteristics of our model, we can stimulate the model with a square wave (pulse) stimulus as shown in \textbf{Figure \ref{fig:pulse-response-ctrnn}A}. The resultant excitatory and inhibitory neural membrane potential responses ($V_e\left(t\right)$ and $V_i\left(t\right)$) are shown in \textbf{Figure \ref{fig:pulse-response-ctrnn}B} and \textbf{Figure \ref{fig:pulse-response-ctrnn}C}. It can be said that, the network has shown both transient and sustained responses. In \textbf{Figure \ref{fig:pulse-response-ctrnn}D}, the excitatory firing rate response $r_e\left(t\right)$ which is related to excitatory potential as $r_e\left(t\right)=g_e\left(V_e\left(t\right)\right)$ is shown. The response $V_i\left(t\right)$ is slightly delayed which leads to the depolarization of excitatory unit until $t=250$ ms. This delay is also responsible from the subsequent re-polarization and plateau formation in the membrane potential of excitatory neuron. The firing rate $r_e\left(t\right)$ is higher during excitation and lower in subsequent plateau and repolarization phases (\textbf{Figure \ref{fig:pulse-response-ctrnn}D}).
\subsection{Spike Generation}
\label{sub:spike-gen}
As we have discussed in \textbf{Section \ref{sub:intro-modeling}}, we will not have any measurement of membrane potential $ V_e(t) $ or $ V_i(t) $. Instead, we will record the spike timings of the neuron and try to solve a maximum likelihood estimation of network parameters $ \hat{\theta} $ using the likelihood function in \eqref{eq:joint-likelihood-product}. Because of that, the simulation needs a method to generate the spike timings of the neurons. As we know from \cite{shadlen1994noise} that, the spikes obey an inhomogeneous Poisson distribution, the best way to achieve the spike timings is to perform a simulation of an inhomogeneous Poisson process of which firing rate is given by:
\begin{equation}\label{eq:firing-rate}
r_e(t)=g_{e}\left(V_{e}\right)=\frac{\Gamma_e}{1+\exp\left( -a_{e}\left(V_{e}-h_e\right)\right) }
\end{equation} 
There are numerous methodologies to generate the Poisson events given the event rate $ r_e(t) $. These ranging from discrete simulation \cite{uteden2008pointprocesses} to thinning \cite{lewis1979simulation}. Discrete simulation may be beneficial when one solves the dynamical models by fixed step solvers such as Euler Integration or Runge-Kutta. The only disadvantage of this approach is that, it confines the spikes into discrete time bins. However in a fixed step integration, the situation for different methods is expected to become same.
Discrete simulation of neural spiking can be summarized as shown below:
\begin{enumerate}
	\item Given the firing rate of any neuron as $ r(t) $
	\item Find the probability of firing at time $ t_i $ by evaluating $ p_i=r(t_i)\Delta t $ where $ \Delta t $ is the integration interval. It should be as small as 1 ms. 
	\item Compute a random variable by drawing a sample from a distribution which is uniform between 0 and 1. Define this as $ x_{rand}=U[0,1] $ where $ U $ stands for uniform distribution. 
	\item If $ p_i>x_{rand} $ fire a spike at $ t=t_i $, else do nothing.
	\item Collect spikes as $ S=[t_1,\ldots,t_{N_s}] $ where $ N_s $ will be the number of spikes obtained at a single run of simulation. 
\end{enumerate}  

\subsection{Step-by-step description of the Problem and Simulation}
\label{sub:example-problem-description}        
The working principles in the example problem can be described in a step-by-step fashion as shown below:
\begin{enumerate}
	\item A single run of simulation will last for $ T_f=3 $ seconds. 
	\item The neuron model in \eqref{eq:our-model-matrix-form} will be simulated at the true value of parameters which are given in \textbf{Table \ref{tab:true-values-parameters}} and \textbf{Table \ref{tab:sigmoid-values-ge-gi}} and firing rate data is stored as $ r_k(t) $ where $ k $ is the current number of simulation. 
	\item Firing rate data $ r_m(t) $ is used to generate neural spikes $ S_m $ in the $ m^\text{th} $ run using the methodology defined in \textbf{Section \ref{sub:spike-gen}}. This data will be used to compute the likelihood. The number of spikes will be $ K_m $ at the $ m^\text{th} $ run.  
	\item \label{it:spike-complete} Repeat the simulation $ M $ times to obtain enough number of spikes. 
	\item The spiking data needed by \eqref{eq:complete-likelihood-compact} will be obtained at the \ref{it:spike-complete}\textsuperscript{th} step. However, the firing rate component of \eqref{eq:complete-likelihood-compact} should be computed at the current iteration of the optimization. 
	\item Run an optimization algorithm of which objective computes the firing rate at the current iterated value of the parameters but the spikes from \textbf{Step \ref{it:spike-complete}}.      
\end{enumerate}
\subsection{Optimization Algorithm}
\label{sub:opt-alg}
Theoretically, any optimization algorithm ranging from gradient descent to derivative free simulated annealing. Most of these algorithms are provided as ready made routines in the optimization and global optimization toolboxes of MATLAB. Regardless of the type of algorithm, all of the methods converge to a local optimum and requires an initial guess. As a result, one needs to start from multiple initial guesses to have a adequate amount of local optimum that will allow us to detect the global one. If we have a convex problem, different initial guesses are expected to converge to same local optimum. In this case, our job will be much easier. The main criteria on the choice of the algorithms is the speed of convergence. Though we have a HPC computing facility we should choose the fastest algorithm as we need to collect a huge amount of data to conclude about the efficiency of the project. Some initial evaluations, suggested that most suitable one in this sense is the local optimizer \texttt{fmincon} provided my MATLAB optimization toolbox. The algorithm needed gradient information but it can be provided by itself through finite difference approximations. There will be 14 (this number equals to the number of cores in a local machine) initial guesses and each initial run will be performed on one core. The whole optimization will be run parallel by the \texttt{parfor} parallel for loop structure of MATLAB. The initial guesses are generated randomly from a uniform distribution. 
\subsection{Simulation data}
\label{sub:sim-dat}
The nominal data in the current problem are given in \textbf{Table \ref{tab:sim-data}}. In order to reveal the effect of different number of stimulus components $ N_U $, amplitude level $ A_n $ and number of trials $ N_{it} $ we will repeat the problem for a set of different values of those parameters. The different values of those parameters are provided in \textbf{Table \ref{tab:sim-data-varied}}.   
  
\begin{table}[H] 
	\centering
	\scriptsize
	\caption{The data related to the current problem. The numerical and methodological data presented here are associated with the main simulation which provides the best outcome. The other situations are presented in \textbf{Table \ref{tab:sim-data-varied}}}
	\begin{tabular}{ccc}
		\toprule 
		Parameter & Symbol & Value \\ 
		\midrule
		Simulation Time & $ T_f $ & 3 sec.\\ 
		\midrule
		Number of Trials & $ M $ & 100 \\
		\midrule
		\# of Components in Stimulus & $ N_U $ & 5 \\
		\midrule
		Method of Optimization & N/A & Interior-Point Gradient Descent (MATLAB) \\
		\midrule
		\# of True Parameters & Size($ \theta $) & 8 \\
		\midrule
		Stimulus Amplitude & $ A_n $ & 100 \\
		\midrule 
		Base Frequency & $ f_0 $ & 3.333 Hz \\
		\bottomrule
	\end{tabular} 
	\label{tab:sim-data}
\end{table} 

\begin{table}[H] 
	\centering
	\scriptsize
	\caption{The data related to the analysis of the problem for different number of trials $ N_{it} $, number of stimulus components $ N_U $, stimulus amplitude $ A_n $}
	\begin{tabular}{ccc}
		\toprule 
		Parameter & Symbol & Value(s) \\ 
		\midrule
		Number of Trials & $ M $ & 25, 50, 100 \\
		\midrule
		\# of Components in Stimulus & $ N_U $ & 5, 10, 20 \\
		\midrule
		Stimulus Amplitude & $ A_n $ & 25, 50, 100 \\
		\bottomrule
	\end{tabular} 
	\label{tab:sim-data-varied}
\end{table}
The initial levels of membrane potentials of excitatory and inhibitory neurons are $ V_e(0)=0 $ and $ V_i(0)=0 $. As we will most probably not know the true values of those conditions assumption of zero values should be sufficient. We will repeat the simulation 20 times for each case, so that we will have sufficient number of results to perform a statistical analysis. 
\subsection{Presentation of the numerical results}
\label{sec:numerical-results}
In this section, the numerical results of the maximum likelihood estimation of the parameters of our neuron model in \eqref{eq:our-model-matrix-form} using maximum likelihood estimation through the maximization of \eqref{eq:complete-likelihood-compact} against parameters in \eqref{eq:theta-ctrnn-param}. The optimization is performed using the gradient based interior-point method provided by MATLAB's \texttt{fmincon} algorithm. All the cases in \textbf{Table \ref{tab:sim-data-varied}} are examined under the conditions \textbf{Table \ref{tab:sim-data}}. The overall results are presented in \textbf{Table \ref{tab:results-mean-values}} and \textbf{Table \ref{tab:results-errors}}. The former presents mean values of the estimated parameters (average of the results obtained from 20 runs) and the latter presents the percent errors between each estimated and true parameters respectively. The second table also presents the mean square estimation errors.     
\begin{sidewaystable}[h!]
	\centering
	\caption{Results related to the mean values of parameters $ \theta= \{\theta_i\}=\left[\beta_e,\beta_i,w_e,w_i,w_{ee},w_{ei},w_{ie},w_{ii}\right] $ for the cases in \textbf{Table \ref{tab:sim-data-varied}} under the conditions \textbf{Table \ref{tab:sim-data}}. The estimation is obtained by performing a maximum likelihood estimation algorithm. The model is defined in \eqref{eq:our-model-matrix-form} and the likelihood function is given in \eqref{eq:complete-likelihood-compact}. The definition of the parameters are as follows. $ M $: The number of samples obtained from experiment or simulation with true parameters, $ A_n $: Value of the fixed amplitude parameter, $ N_U $: The number of components in the stimulus $ I $, $ \hat{\beta}_e,\hat{\beta}_i,\hat{w}_e,\hat{w}_i,\hat{w}_{ee},\hat{w}_{ei},\hat{w}_{ie},\hat{w}_{ii} $: the estimated mean value of the parameters.}
	\label{tab:results-mean-values}
	\vspace{8pt}
	\footnotesize
	\begin{tabular}{ccccccccccc}
		\toprule
		$ M $ & $ A_n $ & $ N_U $ & $ \hat{\beta}_e $ & $ \hat{\beta}_i $ & $ \hat{w}_e $ & $ \hat{w}_i $ & $ \hat{w}_{ee} $ & $ \hat{w}_{ei} $ & $ \hat{w}_{ie} $ & $ \hat{w}_{ii} $ \\
		\midrule
		25 & 25 & 5 & 50.784963 & 26.462242 & 1.044166 &  0.869521 & 1.304249 & 1.948451 & 1.036686 & 0.354959 \\
		25 & 25 & 10 & 49.236254 & 23.352694 & 1.054611 & 0.874023 & 1.141083 & 2.062659 & 0.670130 & 0.597159 \\
		25 & 25 & 20 & 52.966907 & 20.920503 & 0.991920 & 0.986881 & 1.092089 & 1.892108 & 0.782530 & 0.440124 \\
		25 & 50 & 5 & 50.178905 & 22.533484 & 1.012319 & 0.738897 & 1.218979 & 2.212285 & 0.806314 & 0.698271 \\
		25 & 50 & 10 & 51.114765 & 23.003088 & 0.992105 & 0.677112 & 1.173898 & 2.092254 & 0.818681 & 0.641171 \\
		25 & 50 & 20 & 52.839929 & 21.848605 & 0.969480 & 0.693631 & 1.168215 & 2.197068 & 0.729356 & 0.681968 \\
		25 & 100 & 5 & 50.279230 & 24.921546 & 1.002309 & 0.680838 & 1.269726 & 2.225213 & 0.732178 & 0.548961 \\
		25 & 100 & 10 & 50.479445 & 25.050555 & 0.996755 & 0.683263 & 1.240134 & 2.139403 & 0.753446 & 0.502532 \\
		25 & 100 & 20 & 50.316877 & 25.556332 & 0.999808 & 0.766084 & 1.255893 & 2.088012 & 0.723169 & 0.432544 \\
		50 & 25 & 5 & 52.219205 & 25.022006 & 0.994682 & 0.742063 & 1.265516 & 1.935378 & 1.008010 & 0.396907 \\
		50 & 25 & 10 & 49.725766 & 20.681170 & 1.052546 & 0.923283 & 1.127695 & 2.104607 & 0.748163 & 0.635595 \\
		50 & 25 & 20 & 51.324923 & 21.539897 & 0.995424 & 0.893915 & 1.107580 & 1.956125 & 0.767280 & 0.481175 \\
		50 & 50 & 5 & 50.169376 & 24.101709 & 1.005459 & 0.691877 & 1.227451 & 2.148370 & 0.813406 & 0.621339 \\
		50 & 50 & 10 & 50.439033 & 23.799402 & 0.999524 & 0.673302 & 1.160842 & 2.043794 & 0.708546 & 0.467480 \\
		50 & 50 & 20 & 51.680818 & 22.993704 & 0.977620 & 0.682414 & 1.172919 & 2.039575 & 0.743141 & 0.487902 \\
		50 & 100 & 5 & 49.743882 & 25.250105 & 1.016202 & 0.688725 & 1.245729 & 2.164003 & 0.719417 & 0.506157 \\
		50 & 100 & 10 & 50.217624 & 24.699096 & 1.001169 & 0.694363 & 1.205991 & 2.052419 & 0.768446 & 0.519312 \\
		50 & 100 & 20 & 50.113511 & 25.672295 & 1.004083 & 0.729465 & 1.254407 & 2.110134 & 0.663150 & 0.395492 \\
		100 & 25 & 5 & 53.420568 & 21.107605 & 0.956667 & 0.747682 & 1.141893 & 1.970979 & 0.856150 & 0.431254 \\
		100 & 25 & 10 & 49.778298 & 25.010756 & 1.010488 & 0.684748 & 1.227044 & 2.103959 & 0.740378 & 0.505004 \\
		100 & 25 & 20 & 49.168693 & 23.693281 & 1.015881 & 0.840155 & 1.138957 & 1.972214 & 0.734933 & 0.441275 \\
		100 & 50 & 5 & 50.093192 & 24.286094 & 0.994554 & 0.674560 & 1.209774 & 2.093595 & 0.783855 & 0.546576 \\
		100 & 50 & 10 & 49.778298 & 25.010756 & 1.010488 & 0.684748 & 1.227044 & 2.103959 & 0.740378 & 0.505004 \\
		100 & 50 & 20 & 49.778298 & 25.010756 & 1.010488 & 0.684748 & 1.227044 & 2.103959 & 0.740378 & 0.505004 \\
		100 & 100 & 5 & 49.778298 & 25.010756 & 1.010488 & 0.684748 & 1.227044 & 2.103959 & 0.740378 & 0.505004 \\
		100 & 100 & 10 & 49.794365 & 23.968427 & 1.002943 & 0.725149 & 1.176164 & 1.983678 & 0.765260 & 0.518333 \\
		100 & 100 & 20 & 50.043779 & 24.983769 & 0.999714 & 0.714721 & 1.220224 & 2.042329 & 0.691130 & 0.418253 \\ 
		& & & & & & & & & & \\
	\end{tabular}
\end{sidewaystable} 

\begin{sidewaystable}[h!]
	\centering
	\caption{Error analysis of the results obtained in \textbf{Table \ref{tab:results-mean-values}}. The parameters are $ \theta= \{\theta_i\}=\left[\beta_e,\beta_i,w_e,w_i,w_{ee},w_{ei},w_{ie},w_{ii}\right] $. The parameters specific to this table other than  $ M $, $ A_n $ and $ N_U $ (these are given in \textbf{Table \ref{tab:results-mean-values}}) are as follows. $ MSE=E[(\theta-\hat{\theta})^T(\theta-\hat{\theta})] $: Mean square error, $ MSEN=E[(1-\frac{\hat{\theta}_i}{\theta_i})^T(1-\frac{\hat{\theta}_i}{\theta_i})] $: Normalized mean square error, the error associated with each parameter $ \theta_i-\hat{\theta}_i $ is normalized by its true value $ \theta $, $ e_i\%=100\frac{|\theta_i-\hat{\theta}_i|}{\theta_i} $: is the percent error associated with each parameter $ \theta_i $.}
	\label{tab:results-errors}
	\vspace{8pt}
	\footnotesize
	\begin{tabular}{ccccccccccccc}
		
		\toprule
		$ M $ & $ A_n $ & $ N_U $ & $ e_1\% $ & $ e_2\% $ & $ e_3\% $ & $ e_4\% $ & $ e_5\% $ & $ e_6\% $ & $ e_7\% $ & $ e_8\% $ & $ MSE $ & $ MSEN $ \\
		\midrule
		25 & 25 & 5 & 1.569927 & 5.848966 & 4.416587 & 24.217342 & 8.687451 & 2.577469 & 48.098040 & 11.260206 & 14.587511 & 0.291750 \\
		25 & 25 & 10 & 1.527491 & 6.589226 & 5.461057 & 24.860372 & 4.909726 & 3.132957 & 4.267179 & 49.289657 & 10.630085 & 0.212602 \\
		25 & 25 & 20 & 5.933814 & 16.317987 & 0.807969 & 40.983045 & 8.992602 & 5.394593 & 11.789980 & 10.031028 & 16.660514 & 0.333210 \\
		25 & 50 & 5 & 0.357810 & 9.866064 & 1.231931 & 5.556738 & 1.581572 & 10.614260 & 15.187758 & 74.567669 & 5.806786 & 0.116136 \\
		25 & 50 & 10 & 2.229531 & 7.987648 & 0.789487 & 3.269749 & 2.175199 & 4.612676 & 16.954429 & 60.292799 & 6.150238 & 0.123005 \\
		25 & 50 & 20 & 5.679859 & 12.605581 & 3.051974 & 0.909792 & 2.648746 & 9.853386 & 4.193687 & 70.491901 & 6.785449 & 0.135709 \\
		25 & 100 & 5 & 0.558460 & 0.313816 & 0.230857 & 2.737388 & 5.810493 & 11.260628 & 4.596794 & 37.240200 & 4.836198 & 0.096724 \\
		25 & 100 & 10 & 0.958891 & 0.202220 & 0.324517 & 2.390959 & 3.344533 & 6.970135 & 7.635074 & 25.632876 & 5.515861 & 0.110317 \\
		25 & 100 & 20 & 0.633755 & 2.225329 & 0.019229 & 9.440596 & 4.657762 & 4.400595 & 3.309917 & 8.136100 & 3.904414 & 0.078088 \\
		50 & 25 & 5 & 4.438410 & 0.088023 & 0.531762 & 6.009031 & 5.459652 & 3.231079 & 44.001381 & 0.773312 & 12.022560 & 0.240451 \\
		50 & 25 & 10 & 0.548468 & 17.275320 & 5.254616 & 31.897549 & 6.025413 & 5.230345 & 6.880490 & 58.898771 & 11.499223 & 0.229984 \\
		50 & 25 & 20 & 2.649847 & 13.840413 & 0.457642 & 27.702181 & 7.701705 & 2.193757 & 9.611463 & 20.293830 & 14.019862 & 0.280397 \\
		50 & 50 & 5 & 0.338751 & 3.593163 & 0.545864 & 1.160441 & 2.287578 & 7.418491 & 16.200864 & 55.334858 & 3.962367 & 0.079247 \\
		50 & 50 & 10 & 0.878067 & 4.802393 & 0.047638 & 3.813953 & 3.263147 & 2.189687 & 1.220805 & 16.870122 & 4.528800 & 0.090576 \\
		50 & 50 & 20 & 3.361636 & 8.025184 & 2.238008 & 2.512324 & 2.256717 & 1.978726 & 6.162982 & 21.975459 & 5.134585 & 0.102692 \\
		50 & 100 & 5 & 0.512236 & 1.000421 & 1.620176 & 1.610754 & 3.810758 & 8.200145 & 2.773809 & 26.539372 & 3.029622 & 0.060592 \\
		50 & 100 & 10 & 0.435249 & 1.203616 & 0.116876 & 0.805309 & 0.499216 & 2.620952 & 9.778010 & 29.828036 & 3.652699 & 0.073054 \\
		50 & 100 & 20 & 0.227021 & 2.689182 & 0.408299 & 4.209231 & 4.533948 & 5.506697 & 5.264306 & 1.127033 & 3.136577 & 0.062732 \\
		100 & 25 & 5 & 6.841136 & 15.569580 & 4.333293 & 6.811669 & 4.842221 & 1.451062 & 22.307106 & 7.813415 & 10.868654 & 0.217373 \\
		100 & 25 & 10 & 0.443404 & 0.043024 & 1.048782 & 2.178804 & 2.253654 & 5.197952 & 5.768260 & 26.251094 & 2.181165 & 0.043623 \\
		100 & 25 & 20 & 1.662613 & 5.226876 & 1.588053 & 20.022176 & 5.086879 & 1.389302 & 4.990451 & 10.318791 & 11.040645 & 0.220813 \\
		100 & 50 & 5 & 0.186384 & 2.855622 & 0.544644 & 3.634265 & 0.814504 & 4.679771 & 11.979294 & 36.644033 & 3.391140 & 0.067823 \\
		100 & 50 & 10 & 0.443404 & 0.043024 & 1.048782 & 2.178804 & 2.253654 & 5.197952 & 5.768260 & 26.251094 & 2.181165 & 0.043623 \\
		100 & 50 & 20 & 0.443404 & 0.043024 & 1.048782 & 2.178804 & 2.253654 & 5.197952 & 5.768260 & 26.251094 & 2.181165 & 0.043623 \\
		100 & 100 & 5 & 0.443404 & 0.043024 & 1.048782 & 2.178804 & 2.253654 & 5.197952 & 5.768260 & 26.251094 & 2.181165 & 0.043623 \\
		100 & 100 & 10 & 0.411269 & 4.126293 & 0.294329 & 3.592774 & 1.986369 & 0.816080 & 9.322831 & 29.583239 & 2.196393 & 0.043928 \\
		100 & 100 & 20 & 0.087559 & 0.064925 & 0.028608 & 2.102997 & 1.685304 & 2.116475 & 1.267138 & 4.563257 & 2.059480 & 0.041190 \\
		\bottomrule
	\end{tabular}
\end{sidewaystable}
\section{Discussion \& Conclusion}
\subsection{Summary \& General Discussion}
\label{sec:sum-gen-disc}
This research is a devoted to a theoretical study of model fitting to noisy stimulus/response data obtained from sensory neurons. Sensory neurons are known to code the transmitted information in the temporal position of the peaks of their generated successive action potentials. It is also known that, the temporal distribution of the peaks obey inhomogeneous Poisson process where the event rate is considered as a neural firing rate. This firing characteristic allows us to implement a maximum likelihood estimation of the parameters of the fitted model. We use a likelihood function derived from local Bernoulli process which is a function of both the firing rate and the location of individual spikes. The stimulus is modeled as a real phased cosine Fourier series fixed amplitude and frequency but random phase. The maximization of the likelihood is performed by gradient based interior-point method (available as \texttt{fmincon} function in MATLAB). 
\subsection{Evaluation of the Results}
\label{sec:eval-result}
The main results of this research are available in \textbf{Tables \ref{tab:results-mean-values}} and \textbf{\ref{tab:results-errors}}. The first table shows the mean values of the estimated parameters against varying values of the number of samples $ N_{it} $, amplitude $ A_n $ and stimulus order $ N_U $. The second table makes a similar presentation but it has the mean square and percent errors between the estimated and true parameters. According to these results one can make the following comments:
\begin{enumerate}
	\item The main actors that affect the mean square errors of estimation appeared to be the number of samples $ N_{it} $ collected from experiment or simulation with true parameters and the level of stimulus amplitude $ A_n $.
	\item The mean square errors does not show a considerable variation with the number of components in stimulus $ N_U $. 
	\item The individual percentage errors revealed that the number of stimulus components $ N_U $ has an effect on the relative level of the errors. However, this seemed to be more apparent when $ N_{it} $ and $ A_n $ are large. 
	\item Among all these, the best result is shown to be given by $ N_{it}=100 $, $ A_n=100 $ and $ N_U=20 $.  
\end{enumerate}  
\subsection{Future Work}
This study is a fairly new contribution to the theoretical neuroscience literature. Thus, there are a few points that can be addressed in future studies. These may be:
\begin{enumerate}
	\item Application of different fundamental frequencies $ f_0 $ and overall simulation time $ T_f $. 
	\item A different stimulus profies can be applied. These may be pure noise, exponential function, ramp or parabola. 
	\item An interesting application on the same model is to derive the stimulus through an optimal design process. At least the amplitude and frequency component can be optimally calculated using information maximization approaches. Generally Fisher Information Metric is the main objective function here. An approach is given in \cite{doruk2016adaptive}.  
\end{enumerate}     
\section*{Compliance with Ethical Standards}
\subsection*{Funding} This study was partially supported by Turkish Scientific and Technological Research Council's DB-2219 Grant Program. 
\subsection*{Conflict of Interest} The authors declare that they have no conflict of interest. 
\section*{References}
%\bibliographystyle{elsarticle-num} 
%\bibliography{../rdkzpaper}

\end{document}